# MEASURING THE PERIOD OF THE δ SCUTI VARIABLE U1425-01208594 IN CASSIOPEIA

Jeremy Shears [1], David Boyd [2], Steve Brady [3], Ian Miller [4], Roger Pickard [5]

1) British Astronomical Association (BAA) Variable Star Section, "Pemberton", School Lane, Bunbury, Tarporley, Cheshire, CW6 9NR, UK, bunburyobservatory@hotmail.com
2) British Astronomical Association (BAA) Variable Star Section, 5 Silver Lane, West Challow, Wantage, Oxon, OX12 9TX, UK, drsboyd@dsl.pipex.com
3) American Association of Variable Star Observers (AAVSO), 5 Melba Drive, Hudson, NH 03051, USA, sbrady10@verizon.net
4) British Astronomical Association (BAA) Variable Star Section, Furzehill House, Ilston, Swansea, SA2 7LE, UK, furzehillobservatory@hotmail.com
5) British Astronomical Association (BAA) Variable Star Section, 3 The Birches, Shobdon, Leominster, Herefordshire, HR6 9NG, UK, rdp@astronomy.freeserve.co.uk

**Abstract:** The variability of U1425-01208594 was recently discovered by Schmidtobreick *et al* (2003), who suggested that it is a member of the δ Scuti family of pulsating stars. Photometry conducted by the authors revealed a period of 0.06695(8) d and a peak-to-peak amplitude of 0.014 mag.

**Introduction**

The variability of U1425-01208594 (USNO-A2.0 catalogue, RA 00 51 58.150 Dec. +53 51 40.93) was recently discovered by Schmidtobreick *et al.* (2003) during studies on the nearby cataclysmic variable V452 Cas. They found the star varied with a peak-to-peak amplitude of ~0.02 mag. and a period of 0.060(2) d, although this was only based on two nights of photometry covering a total of ~4 cycles. Further spectroscopic studies revealed the object to be of spectral type A3 to A7. The spectral classification, short period and small amplitude led them to conclude that the object is most likely a δ Scuti type variable. δ Sct variables are pulsating stars located in the lower part of the Cepheid instability strip having short periods (<0.3 d) and visual amplitudes ranging from a few thousands of a magnitude to several tenths, with typical amplitudes of 0.02 mag (Rodriguez *et al.* 2000, Rodriguez and Breger 2001).

**Time resolved photometry**

During 2007 September, the authors conducted unfiltered time resolved photometry on V452 Cas during its superoutburst. Since the new δ Sct variable was located only 3 arcmin away, they were also able to extract photometry on this star from the same images; these data were supplemented by a further run in October. Table 1 summarises the instrumentation used and Table 2 contains a log of the time-series runs. Exposure times of 60 s were optimised for photometry of V452 Cas, which is considerably fainter than U1425-01208594. In all cases raw images were flat-fielded and dark-subtracted, before being analysed using commercially available aperture photometry software against the comparison star sequence given in the AAVSO chart for V452 Cas dated 020301. JS used stars 144 and 154 as comparison and check respectively. DB, IM and RP used a comparison star ensemble comprising stars 144, 146, 150 and 154, whereas SB used an ensemble of 144, 154, 156, 158 and 166. An image of the field identifying the variable is shown in Figure 1.

| Observer | Telescope | CCD |
|---|---|---|
| JS | 0.28 m SCT | Starlight Xpress SXV-M7 |
| DB | 0.35 m SCT | Starlight Xpress SXV-H9 |
| SB | 0.4 m reflector | SBIG ST-8XME |
| IM | 0.35 m SCT | Starlight Xpress SXVF-H16 |
| RP | 0.30 m SCT | Starlight Xpress SXV-H9 |

Table 1: Equipment used

| Date in 2007 (UT) | Start time (JD-2454000) | Duration (h) | Observer |
|---|---|---|---|
| Sep 2 | 345.564 | 7.3 | SB |
| Sep 3 | 346.566 | 7.4 | SB |
| Sep 3 | 347.345 | 4.0 | JS |
| Sep 3 | 347.364 | 4.7 | RP |
| Sep 3 | 347.383 | 3.3 | DB |
| Sep 4 | 348.343 | 4.2 | IM |
| Sep 4 | 348.354 | 2.4 | RP |
| Sep 6 | 350.330 | 3.4 | JS |
| Sep 6 | 350.392 | 3.5 | RP |
| Sep 7 | 351.337 | 2.1 | JS |
| Sep 7 | 351.359 | 7.2 | RP |
| Sep 7 | 351.359 | 7.1 | IM |
| Sep 10 | 354.334 | 4.2 | JS |
| Sep 10 | 354.344 | 4.2 | DB |
| Sep 10 | 354.400 | 6.9 | IM |
| Sep 10 | 354.415 | 4.0 | RP |
| Sep 11 | 355.389 | 2.0 | RP |
| Sep 12 | 356.359 | 4.2 | DB |
| Sep 12 | 356.403 | 4.5 | RP |
| Sep 15 | 358.564 | 2.9 | IM |
| Oct 29 | 402.779 | 2.7 | SB |

Table 2: Log of time-series observations

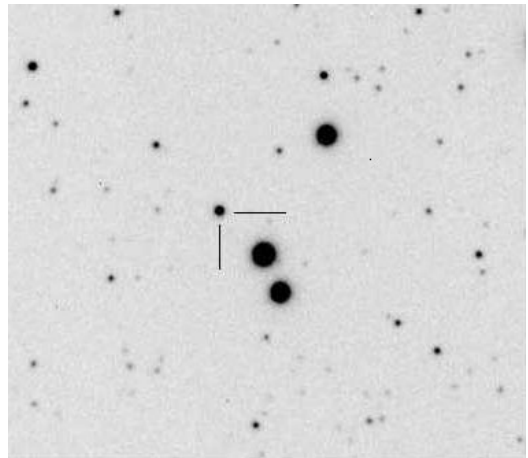

Figure 1: Field of U1425-01208594 (marked) imaged on 2007 Sep 3.44
Field approximately 5' x 5' with S at top and E to right

**Period determination**

To determine the period of the brightness variation, we used the Lomb-Scargle algorithm (Lomb 1976, Scargle 1982) as implemented in the Peranso software (Vanmunster 2007), having normalized the data by subtracting the arithmetic mean of all the magnitudes. The resulting power spectrum is shown in Figure 2. The largest peak in the spectrum corresponds to a period of 0.06695(8) d. The next two largest peaks, at 0.06276(7) d and 0.07173(9) d, are the 1 c/d alias of the major peak. Pre-whitening the power spectrum with the predominant period P= 0.06695 d leaves no significant residual period (Figure 3).

A phase diagram of the data, folded on the P= 0.06695 d, is shown in Figure 4. The peak-to-peak amplitude is 0.014 mag.

The period analysis was also repeated using the ANOVA method (Schwarzenberg-Czerny 1996) as implemented in Peranso. This gave results which are fully consistent with the above analysis.

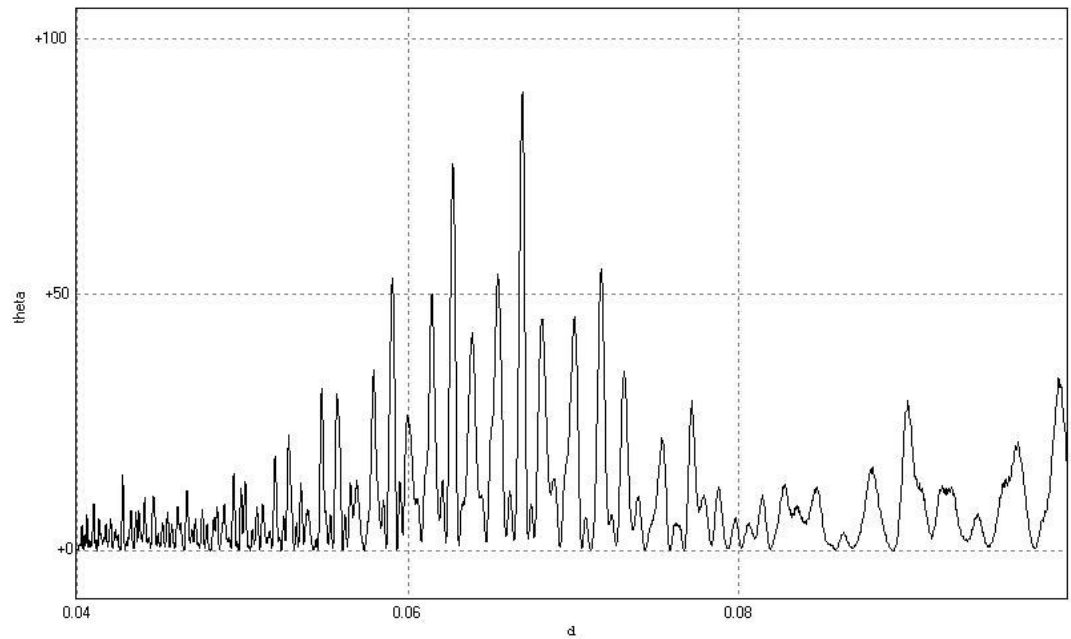

Figure 2: Power spectrum of combined data

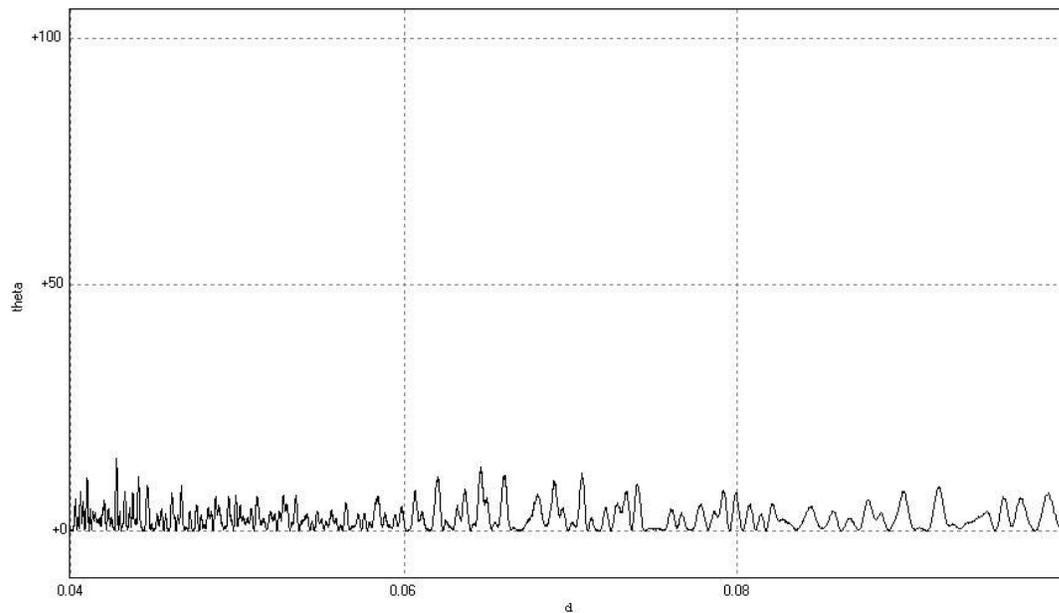

Figure 3: Power spectrum of the data shown in Figure 2 after pre-whitening with the period P= 0.06695 d

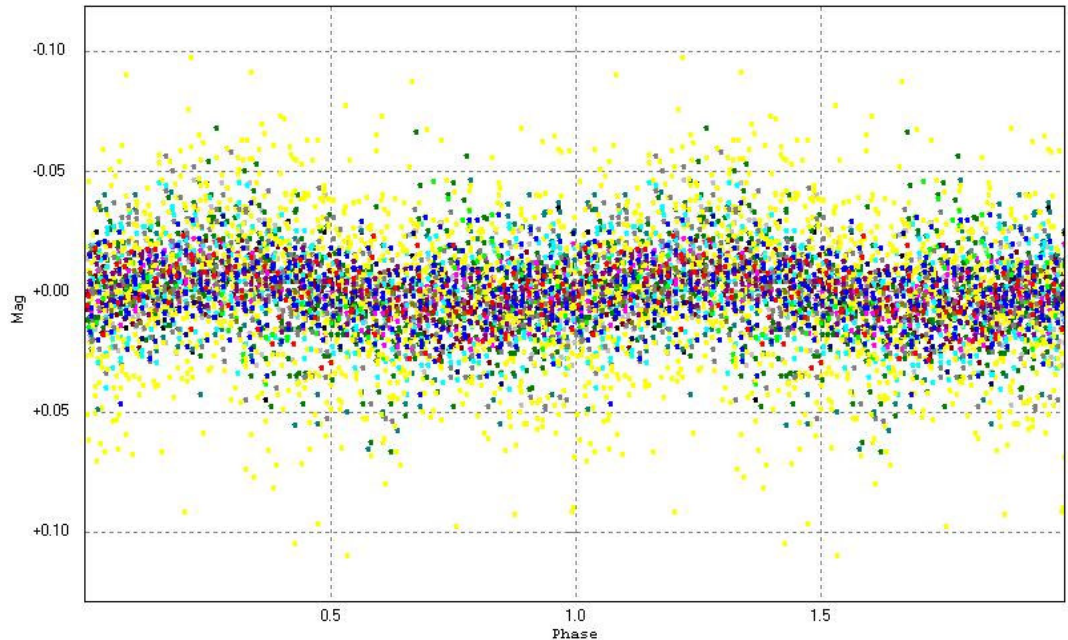

Figure 4: Phase diagram of data folded on P= 0.06695 d

**Conclusion**

We found that U1425-01208594 has a period of 0.06695(8) d and a peak-to-peak amplitude of 0.014 mag. Our data support the assertion of Schmidtobreick *et al.* (2003) that this star is a δ Sct variable, although our period is slightly longer than the period measured by them. We suggest that further photometry would be of value to look for multi-periodic variations such as have been seen in some δ Sct stars.